\documentclass[showpacs,pre]{revtex4}
\usepackage{amsmath}
\usepackage{graphicx}
\begin{document}
\title{Mechanism of Stepped Leaders in a Simple Discharge Model}
\author{Hidetsugu Sakaguchi and Sahim M. Kourkouss}
\affiliation{Department of Applied Science for Electronics and Materials,
Interdisciplinary Graduate School of Engineering Sciences, Kyushu
University, Kasuga, Fukuoka 816-8580, Japan}
\begin{abstract} 
  We construct a one-dimensional model for the stepped leader in the filamental discharge by simplifying  an electric-circuit model of discharge. We find that the leader of the discharge moves stepwise by direct numerical simulations, and then we try to understand the mechanism of the stepwise motion by reducing the spatially extended system to the dynamics of the tip position of the discharge. 
\end{abstract}
\maketitle
\section{Introduction}
  The discharge occurs when the voltage between two electrodes is beyond a critical value. There are various forms of discharge, such as corona discharge, spark discharge, glow discharge, and arc discharge, depending on various conditions, such as pressure, temperature and the shape of electrodes.  
For gas discharge, the pressure $p$ and the gap length $l$ between the electrodes are important.  When $p\times l$ is large, the spark discharge takes a form of a filament. Meek proposed the streamer theory for the filamental discharge.~\cite{rf:1}  The tip region of a growing filamental discharge is sometimes called a leader.
It is known that the leader grows stepwise in the lightning discharge. The gap distance between the electrodes is very long in the lightning discharge.~\cite{rf:2}  The leader goes down from a thunder cloud to the ground relatively slowly. The velocity is around $10^5$ m/s. 
The extending process of the leader is invisible to the eyes, but it can be observed with a high-speed camera. The leader moves in steps of about 30 m with a pause of about 40 ms between steps.  This type of leader is called the stepped leader.  When the leader reaches the ground,  a strong flash called a return stroke appears instantly, which we observe as lightning. However, the origin of the stepped motion is not completely understood. 
There were several qualitative theories for the stepped leaders. 
Bruce pointed out the importance of the transition from a weak glow discharge to a strong arc discharge.~\cite{rf:3} Kumar and Nagabhushana proposed a simulation model of stepped leaders based on a complex electric breakdown process.~\cite{rf:4}  
In a previous study, we proposed a simple deterministic electric circuit model composed of resistors and capacitors, and performed a numerical simulation on triangular lattices to investigate complex patterns in discharge processes.~\cite{rf:5} In the model, a two-step function for the conductance is assumed, which expresses a transition from a weak discharge to a strong discharge.  Branched patterns of the discharge similar to the Lichtenberg figure were found in the numerical simulation of the model. We found a stepwise motion of the leaders in a certain parameter range and showed that the branching of the pattern is not directly related to the stepped motion. In this paper, we simplify the electric circuit model to a one-dimensional model and try to understand a mechanism of the stepped motion qualitatively. 

\section{One-Dimensional Model and Numerical Simulation}
We consider a one-dimensional electric circuit model in this paper to simplify the argument.  The gap length between the electrodes is assumed to be $L$, and the voltage $V_0$ is applied between $x=0$ and $x=L$. In the numerical simulation, the space of size $L$ is discretized with the interval $\Delta x$. The interval $\Delta x=0.1$ is used in most numerical simulations. 
The electric potential at the $i$th site is denoted by $V_i$ at $i=x/\Delta x$. A resistor is set between the $i$th site and the $(i+1)$th site, and the conductance is expressed as $\sigma_i/\Delta x$. A capacitor is set between the $i$th site and the earth, and the capacitance is assumed to be $C\cdot \Delta x$. Here, we assume that the conductance is inversely proportional to the interval $\Delta x$ and the capacitance is proportional to $\Delta x$. The time evolution of $V_i$ is expressed as
\begin{equation}
C\frac{dV_i}{dt}=\{\sigma_{i-1}(V_{i-1}-V_i)-\sigma_i(V_i-V_{i+1})\}/(\Delta x)^2.
\end{equation}
By the continuum approximation, eq.~(1) is reduced to the partial differential equation 
\begin{equation}
C\frac{\partial V}{\partial t}=\frac{\partial}{\partial x}\left ( \sigma(x)\frac{\partial V}{\partial x}\right ),
\end{equation}
where $V(x)$ and $\sigma(x)$ denote $V_i$ and $\sigma_i$ at $i=x/\Delta x$, respectively. 
The boundary conditions for $V(x)$ are expressed as  $V(x)=V_0$ at $x=0$ and $V(x)=0$ at $x=L$.
Similarly to the previous paper, we assume a two-step function for the conductance, although we have observed a similar behavior even if the two-step function is slightly modified to a continuous function.  Thus, the time evolution of $\sigma_i$ is expressed as  
 \begin{eqnarray}
\tau\frac{d\sigma_i}{dt}&=&-\sigma, \;\;\;{\rm for}\;\; E_i<E_{c1},\nonumber\\
\tau\frac{d\sigma_i}{dt}&=&\sigma_1-\sigma, \;\;\;{\rm for}\;\; E_{c1}<E_i<E_{c2},\nonumber\\
\tau\frac{d\sigma_i}{dt}&=&\sigma_2-\sigma, \;\;\;{\rm for}\;\; E_{c2}>E_{i},
\end{eqnarray} 
where $E_i=|V_{i+1}-V_{i}|/\Delta x$ is the local voltage between the $i$th and  $(i+1)$th sites, and $E_{c1}$ and $E_{c2}$ are threshold values for the weak and strong discharges, $\tau$ is the relaxation time, and $\sigma_{1,2}$ are stationary values of the conductance. We assumed that $\sigma_2$ is much larger than $\sigma_1$ because the ionization proceeds rapidly at the transition from the weak discharge to the strong discharge. 
We further assume that the threshold $E_{c1}$ and $E_{c2}$ decrease with $\sigma$ as $E_{c1}=E_{c10}-\alpha \sigma$ and $E_{c2}=E_{c20}-\alpha\sigma$. 
If the local voltage is below the first threshold $E_{c1}$ in the entire region, the conductance decays to 0, which implies the insulator. 
When the local voltage is increased and goes beyond the first threshold, the discharge occurs and the conductance becomes nonzero. Then, the first threshold value slightly decreases as $E_{c1}=E_{c10}-\alpha \sigma$. This effect induces a hysteresis in the conductance.  That is, if the discharge occurs once, the discharged state is maintained even if the local voltage is slightly decreased.  A similar hysteresis is assumed to occur owing to the term $E_{c2}=E_{c20}-\alpha\sigma$  at the transition from the weak discharge to the strong discharge. This type of hysteresis is often observed even in experiments of discharge phenomena.

If $E_{c2}$ is sufficiently large, the strong discharge does not occur, because $E_i$ does not reach the second threshold. In this case, the leader or the tip position of the weakly discharged state moves smoothly. We show a numerical result in Fig.~1. The parameter values are $V_0=1,L=90, \Delta x=0.1,C=0.03,\tau=0.01,\alpha=0.02,\sigma_1=50,\sigma_2=2000,E_{c1}=0.1$, and $E_{c2}=10$, and 
the initial conditions are set to be $\sigma_i=0$ and $V_i=0$.
Figure 1(a) shows the time evolution of the tip position of the weak discharge $x_1=i_1\Delta x$.  The tip position $x_1$ increases monotonically, and the velocity $v_1=dx_1/dt$ around $t=0.08$ is evaluated at $v_1\sim 280$. However, the propagation velocity slightly decreases in time. This is because the distance between the electrode at $i=0$ and the tip position $i_1$ increases in time.  Figures 1(b) and 1(c) show profiles of the conductance $\sigma(z)$ and the electric potential $V(z)$ at $t=0.08$. Here, $z=x-x_1$ denotes the distance from the tip position $x_1$. For $z>0$, $\sigma(z)=0$ and $V(z)=0$, which implies that $z>0$ ($i>i_1$) is an insulator region.  The conductance $\sigma(z)$ increases gradually to $\sigma_1$ as $|z|$ ($z<0)$ is distant from $z=0$. 
 The dashed line in Fig.~1(c) denotes $V(z)=-E_{c10}z$, which implies that the local voltage $E$ at the tip of the discharge is equal to $E_{c10}$. 
Although the local voltage $E=|\partial V/\partial z|$ is smaller than $E_{c10}$ in almost the entire region of $z<0$ except for the region $z\sim 0$, the discharged state is maintained because of the hysteresis effect, that is, $E$ is smaller than $E_{c10}$ but larger than $E_{c1}=E_{c10}-\alpha\sigma(z)$.   

\begin{figure}[t]
\begin{center}
\includegraphics[height=4.cm]{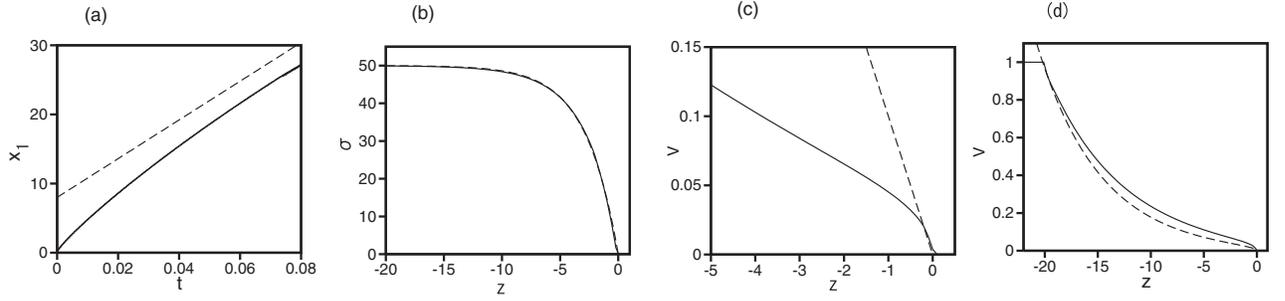}
\end{center}
\caption{(a) Time evolution of the tip position $x_1(t)$ of the discharge for $V_0=1,C=0.03,\Delta x=0.1,\alpha=0.02,\sigma_{1}=50,\sigma_{2}=2000,\tau=0.01,E_{c10}=0.1,E_{c20}=10$, and $L=90$. The dashed line denotes $x_1(t)=v_1t+x_0$ with $v_1=280$. (b) Snapshot profile of $\sigma(z)$ where $z=x-x_1$. The dashed curve denotes the theoretical curve obtained using eq.~(5). (c) Snapshot profile of $V(z)$ in the range of $-5<z<0$. The dashed line denotes $V=-0.1\cdot z$. (d) Snapshot of $V(z)$ in the range of $-22<z<1$. The solid curve denotes the numerical result and the dashed curve denotes eq.~(7). 
}
\label{f1}
\end{figure}
If we assume that the tip position $x_1$ moves at a constant velocity $v_1$, steadily moving solutions $\sigma(z)=\sigma(x-v_1t)$ and $V(z)=V(x-v_1t)$ can be obtained from eqs.~(2) and (3).  Equation (3) leads to 
\begin{equation}
-v_1\tau\frac{\partial \sigma}{\partial z}=\sigma_1-\sigma,
\end{equation}
for $z<0$. The solution to eq.~(4) is expressed as
\begin{equation}
\sigma(z)=\sigma_1[1-\exp\{z/(v_1\tau)\}],
\end{equation}
for $z<0$ and $\sigma(z)=0$ for $z>0$.
The velocity $v_1$ is evaluated at 280 near $t=0.08$ from Fig.~1(a). The dashed curve in Fig.~1(b) denotes eq.~(5) using $v_1=280,\tau=0.01$, and $\sigma_1=50$. Good agreement is seen between the numerical and theoretical curves. On the other hand, eq.~(2) is reduced to 
\begin{equation}
-v_1 C V=\sigma(z)\frac{\partial V}{\partial z}.
\end{equation}
The substitution of eq.~(5) into eq.~(6) yields the solution
\begin{equation} 
V(z)=V(z_0)\exp[(-v_1^2C\tau/\sigma_1)\{(z-z_0)/(v_1\tau)+\ln (1-e^{z_0/(v_1\tau)})-\ln (1-e^{z/(v_1\tau)} )\}],
\end{equation}
where $z_0$ is a certain position where $V(z)=V(z_0)$ is satisfied. 
Figure 1(d) shows a comparison of $V(z)$ (solid curve) obtained by a direct numerical simulation and $V(z)$ (dashed curve) expressed by eq.~(7), where $V(z_0)=V_0=1$ at $z_0=-20.1$ (or $x=0$) is used. 
The deviation of the two curves is considered to originate from the nonstationarity of this system, that is, the boundary condition $V(z_0)=V_0$ at the left electrode $z_0=-v_1t$ is a moving boundary condition in this moving frame.     

\begin{figure}[t]
\begin{center}
\includegraphics[height=4.cm]{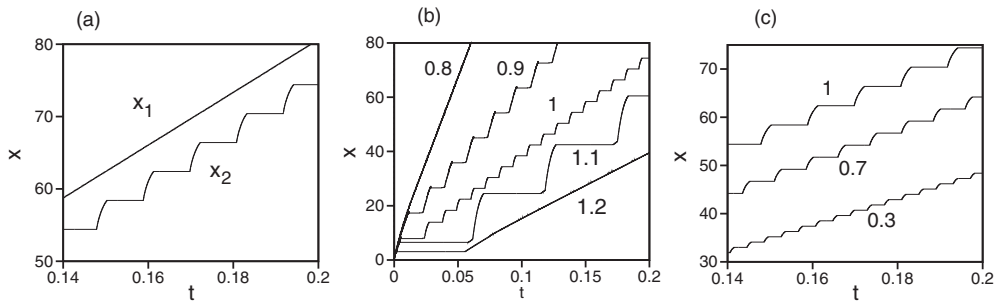}
\end{center}
\caption{(a) Time evolutions of the tip positions $x_{1}(t)$ and $x_2(t)$ of the weak and strong discharges for $V_0=1,\Delta x=0.1, C=0.03,\alpha=0.02,\sigma_{1}=50,\sigma_{2}=2000,\tau=0.01,E_{c10}=0.1,E_{c20}=1$, and $L=90$. (b) Time evolutions of the tip positions $x_{2}(t)$ at $E_{c20}=0.8,0.9,1,1.1$, and 1.2 for $V_0=1$. 
(c) Time evolutions of the tip positions $x_{2}(t)$ at $V_0=0.3,0.7$, and 1 for $E_{c20}=1$. 
}
\label{f2}
\end{figure}
\begin{figure}[t]
\begin{center}
\includegraphics[height=4.cm]{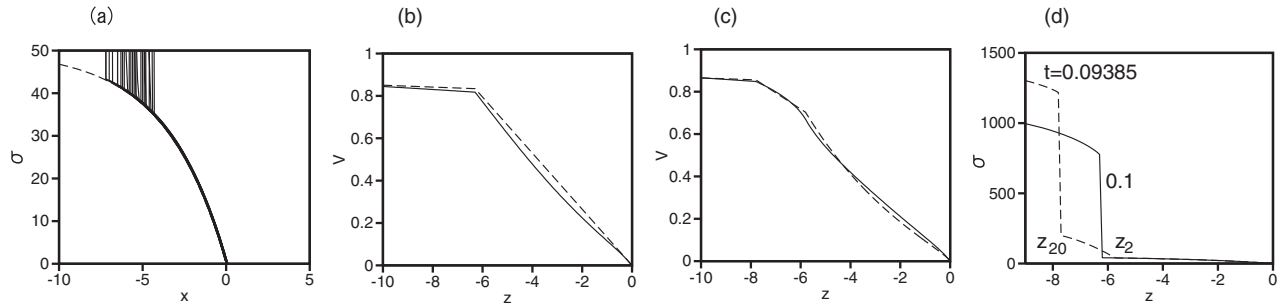}
\end{center}
\caption{ (a) Snapshot profiles of $\sigma(z)$ where $z=x-x_1$. The dashed curve denotes the theoretical curve obtained using eq.~(5). (b) Snapshot profile of $V(z)$ at $t=0.1$ where $z=x-x_1$ in a stepped stage. The dashed curve denotes a piecewise linear function of $x$.  (c) Snapshot profile of $V(z)$ at $t=0.09385$ where $z=x-x_1$ in a moving stage. The dashed curve denotes a linked curve of a piecewise linear function (10) and the function expressed by eq.~(11). (d) Snapshot profiles of $\sigma(z)$ at $t=0.1$ (solid curve) and $t=0.09385$ (dashed curve).}
\label{f3}
\end{figure}

We have performed several numerical simulations by changing the second threshold $E_{c20}$ at a fixed value of $V_0=1$. The other parameter values and the initial conditions are the same as those previously used. We have found that the strong discharge does not appear at $E_{c20}>1.8$. The strong discharge occurs for $E_{c20}<1.8$, and the tip position $x_2$ of the strong discharge moves stepwise for $0.85<E_{c20}<1.17$.  $x_2$ moves smoothly for $E_{c20}<0.85$. 
Figure 2(a) shows time evolutions of the tip positions $x_1$ and $x_2$ respectively for the weak and strong discharges at $E_{c20}=1$ and $V_0=1$. It is clearly seen that $x_2(t)$ moves stepwise, that is, $x_{2}(t)$ repeats the forward motion and the stop. On the other hand, $x_1(t)$ moves smoothly. Figure 2(b) shows time evolutions of $x_2(t)$ for $E_{c20}=0.8,0.9,1,1.1$, and 1.2. The average velocity of $x_2(t)$ decreases with $E_{c20}$. The period of the stepped motion is the shortest near $E_{c20}=1$. 

Several numerical simulations were also performed by changing $V_0$ at a fixed value of $E_{c20}=1$.  Figure 2(c) shows the time evolution of $x_{2}(t)$ for $V_0=0.3,0.7$, and 1. Stepwise motions are clearly observed.  The average velocity of the tip position slightly increases with $V_0$; however, the period of the stepped motion changes rather strongly with $V_0$. 

The stepped motion of the leader has been studied in more detail for $E_{c20}=1$ and $V_0=1$. 
The tip position $x_{2}$ of the strong discharge exhibits a stepped motion, but the tip position $x_{1}$ of the weak discharge moves smoothly, as shown in Fig.~2(a). The velocity of the first tip $x_{1}$ of the weak discharge is evaluated as $v_1=364$ at the parameter values. Figure 3(a) displays several snapshots of $\sigma(z)$ at different times, where $z$ is the distance $z=x-x_{1}$ from the first tip $x_{1}$. The dashed curve is $\sigma(z)=\sigma_1[1-\exp\{z/(v_1\tau)\}]$. In the region of the weak discharge, the stationary solution of $\sigma(z)$ by eq.~(5) is rather good approximation.  When the stronger discharge sets in, $\sigma(z)$ increases rapidly. Almost discontinuously jumped lines in Fig.~3(a) correspond to the transition to the strong discharge. 

Figure 3(b) displays a snapshot pattern of $V(z)$ at $t=0.1$ when the leader is in a stepped stage. The position of $x_{2}$ remains constant with time in the stepped stage. The conductance $\sigma(z)$ at $t=0.1$ is shown in Fig.~3(d) with a solid curve. The conductance is very large, i.e., $\sigma\sim \sigma_2^{\prime}\sim\sigma_2$ for $z<z_{2}=x_2-x_1$, and rather small, i.e., $\sigma \sim \sigma_1^{\prime}\sim \sigma_1$ for $z_2<z<0$. 
If $\sigma_1^{\prime}$ and $\sigma_2^{\prime}$ are assumed to be certain constant values, $V(x)$ can be roughly approximated at a piecewise linear function:
\begin{eqnarray}
V(x)&=&V_0+\frac{(V_2-V_0)x}{x_{2}},\;\;\; {\rm for}\;\;0<x<x_2,\\
&=& \frac{V_2(x_{1}-x)}{x_{1}-x_{2}},\;\;\; {\rm for}\;\;x_2<x<x_1,
\end{eqnarray}
where 
\[V_2=V(x_{2})=\frac{V_0}{1+\sigma_1^{\prime}x_{2}/\{\sigma_2^{\prime}(x_{1}-x_{2})\}}.\]
If $\sigma_1^{\prime}=40$ and $\sigma_2^{\prime}=1200$ are used, $V_2$ is evaluated at $V_2=0.811$ for $x_{2}=37.7$ and $x_{1}=44$ corresponding to the snapshot profile shown in Fig.~3(b). The dashed curve in Fig.~3(b) denotes this piecewise linear approximation of $V(z)$. 

Figure 3(c) displays a snapshot pattern of $V(z)$ at $t=0.09385$ when the leader is in a moving stage.  The second tip $x_{2}$ moves approximately with the velocity $v_2\sim 1640\sim 4.5v_1$. 
The shoulderlike structure is characteristic of the profile of $V(z)$ in contrast to the profile in Fig.~3(b), in which the local voltage $E=|\partial V/\partial z|$ is rather large near $z=z_{2}$.  The conductance $\sigma(z)$ is shown in Fig.~3(d) with a dashed curve. 
The conductance $\sigma(z)$ and the derivative of $V(z)$ have a discontinuity at $z_{20}\sim -7.8$.  Here, the discontinuity point $x_{20}=x_1+z_{20}$ is the position where the tip $x_2$ of the strong discharge remained in the previous stepped stage. 
The conductance in the region of $x_{2}<x<x_{1}$ is approximated at $\sigma_1[1-\exp\{(x-x_{1})/(v_1\tau)\}]$ and the conductance for $x_{20}<x<x_{2}$ is approximated at $\sigma_1[1-\exp\{(x_{2}-x_{1})/(v_1\tau)\}]+(\sigma_2-\sigma_1)[1-\exp\{(x-x_{2})/(v_2\tau)\}]$. 
If the conductance is roughly approximated at certain constant values as $\sigma(x)=\sigma_2^{\prime}$ for $0<x<x_{20}$ and $\sigma(x)=\sigma_1^{\prime}$ for $x_{20}<x<x_{2}$, $V(x)$ is roughly approximated at a piecewise linear function:
\begin{eqnarray}
V(x)&=&V_0+\frac{(V_{20}-V_0)x}{x_{20}}\;\;\; {\rm for}\;\;0<x<x_{20},\nonumber\\
&=&V_{20}+\frac{(V_{20}-V_2)(x_{20}-x)}{x_{2}-x_{20}}\;\;\;{\rm for}\;\;x_{20}<x<x_{2}.
\end{eqnarray}
On the other hand, we use eq.~(7) for $V(x)$ in the region of $x_{2}<x<x_{1}$:
\begin{eqnarray} 
V(x)&=&V(x_2)\exp[(-v_1^2C\tau/\sigma_1)\{(x-x_2)/(v_1\tau)+\ln (1-e^{(x_2-x_1)/(v_1\tau)})-\ln (1-e^{(x-x_1)/(v_1\tau)} )\}] \nonumber\\
&&\hspace{7cm}{\rm for}\;\; x_2<x<x_1.
\end{eqnarray}   
The parameters $V_{20}=V(x_{20})$ and $V_{2}=V(x_{2})$ are unknown. They are determined by the boundary conditions of $V(x)$ at $x=x_{20}$ and $x_{2}$ expressed by 
\begin{equation}
\sigma_2^{\prime}\frac{V_0-V_{20}}{x_{20}}=\sigma_1^{\prime}\frac{V_{20}-V_2}{x_{2}-x_{20}},\;\;\sigma_1^{\prime}\frac{V_{20}-V_2}{x_{2}-x_{20}}=-\sigma(x_2)\frac{\partial V(x_{2})}{\partial x},
\end{equation}
which are due to the continuity of the current. 
The parameters $V_{20}$ and $V_2$ are explicitly expressed as
\begin{eqnarray}
V_{20}&=&\frac{\sigma_2^{\prime}\{\sigma_1^{\prime}+v_1C(x_2-x_{20})\}V_0}{\sigma_2^{\prime}\{\sigma_1^{\prime}+v_1C(x_2-x_{20})\}+\sigma_1^{\prime}v_1Cx_{20}},\nonumber\\
V_2&=&\frac{\sigma_1^{\prime}V_{20}}{\sigma_1^{\prime}+v_1C(x_2-x_{20})}.
\end{eqnarray}
The dashed curve in Fig.~3(c) shows a linked curve of the piecewise linear function (10) and the function expressed by eq.~(11) using $\sigma_2^{\prime}=1800$ and $\sigma_1^{\prime}=100$, although the horizontal axis is shifted to $z=x-x_1$. We have not yet succeeded in obtaining the solution satisfying the correct moving boundary conditions with different velocities $dx_{20}/dt=0,dx_{2}/dt=v_2$, and $dx_{1}/dt=v_1$. However, the above approximate solution reproduces a shoulder structure and a sharp derivative near $x=x_{2}$ fairly well. 

\begin{figure}[t]
\begin{center}
\includegraphics[height=4.cm]{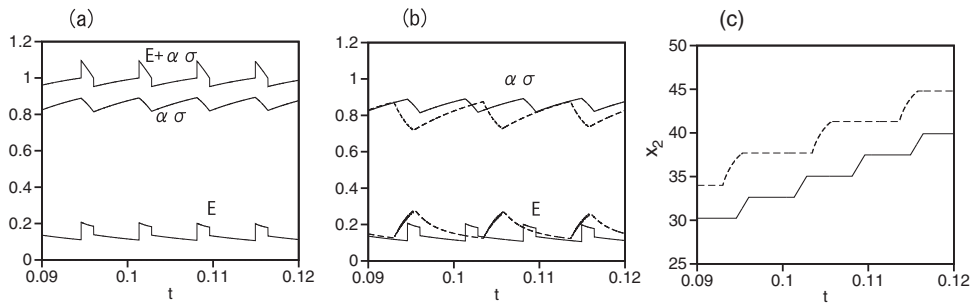}
\end{center}
\caption{(a) Time evolutions of $\alpha\sigma(x_2), E(x_2)$ and $E+\alpha\sigma$ obtained using eqs.~(14)-(17).  (b) Comparison of the time evolutions of $\alpha\sigma(x_2)$ and $E(x_2)$ obtained using eqs.~(14)-(17) (solid curves) and the direct numerical simulation (dashed curves).   
(c) Time evolutions of the tip position $x_{2}(t)$ of the strong discharge. 
The dashed curve denotes the numerical result and the solid one denotes the theoretical approximation. 
}
\label{f4}
\end{figure}
\section{Simple Model for Stepped Leader}
On the basis of these numerical observations and the rough approximation for the profile of the electric potential shown in the previous section, we propose a very simple model for the time evolution of the position $x_2(t)$.  
In the stepped stage, the first tip $x_1$ moves with the velocity $v_1$ as $x_1=x_0+v_1t$, where $x_0$ is a certain initial position, and the second tip $x_2$ remains at a certain position: $x_2(t)=x_{20}$. 
In this stepped stage, the conductance $\sigma(x_2)$ at $x=x_2$ increases as
\begin{equation}
\sigma(x_2)=\sigma_1[1-\exp\{-(x_0+v_1t-x_{20})/(v_1\tau)\}],
\end{equation}
because the difference $x_1-x_2=x_0+v_1t-x_{20}$ increases with time. 
On the other hand, the derivative of the electric potential at $x=x_2$ is evaluated using eq.~(9) as
\begin{equation}
E=\frac{\sigma_2^{\prime}V_0}{\sigma_1^{\prime} x_{20}+\sigma_2^{\prime}(x_{1}-x_{20})},
\end{equation}
where $\sigma_1^{\prime}=40$ and $\sigma_2^{\prime}=1200$ are used in the following numerical simulation. 
In the stepped stage, $E$ decreases monotonically with time, because the difference $x_1-x_2=x_0+v_1t-x_{20}$ increases with time. The time evolutions of $\sigma(x_2),E(x_2)$, and $E(x_2)+\alpha\sigma(x_2)$ are shown in Fig.~4(a). The sum $E+\alpha\sigma(x_2)$ increases monotonically with time and reaches the second threshold $E_{c20}=1$ from below. Namely, the local voltage $E(x_2)$ goes beyond the threshold $E_{c20}-\alpha \sigma(x_2)$.
Then, the gate to the strong discharge opens and the second tip $x_2$ starts to move with the velocity $v_2$.   

In the moving stage, the time evolution of $x_2(t)$ is expressed with $x_2=x_{20}+v_2(t-t_n)$, and the conductance $\sigma(x_2)$ is approximated using eq.~(5) as 
\begin{equation}
\sigma(x_2)=\sigma_1[1-\exp\{-(x_0+v_1t-x_{20}-v_2(t-t_n))/(v_1\tau)\}].
\end{equation}
Here, $t_n$ is the time when a transition to the moving stage occurs.
The conductance $\sigma(x_2)$ decreases monotonically with time in the moving stage because the distance $x_1-x_2=x_0-x_{20}+v_2t_n-(v_2-v_1)t$ decreases with time.   
The derivative $E=\partial V/\partial x$ of the electric potential at $x=x_2$  is evaluated using eq.~(11) as 
\begin{equation}
E=\frac{\sigma_1^{\prime}V_{20}}{\{\sigma_1^{\prime}+v_1C(x_2-x_{20})\}\sigma(x_2)v_1C},
\end{equation}
where the parameter values $\sigma_1^{\prime}=100$ and $\sigma_2^{\prime}=1800$ are used. 
There is a discontinuity in $E$'s of eqs.~(15) and (17)  at $t=t_n$ in this simplified model. The summation of $E(x_2)+\alpha \sigma(x_2)$ decreases monotonically with time and reaches the threshold $E_{c20}=1$ from above at $t=t_n^{\prime}>t_n$. Then, the moving stage changes into the stepped stage because the local voltage $E$ is below the second threshold $E_{c20}-\alpha\sigma$, and $x_2$ stops. The stepped stage continues again until $t=t_{n+1}$. 
The repetition of the stepped and moving stages reproduces the behavior of the stepped leader, as shown in Fig.~4(c). Figure 4(b) shows time evolutions of $\sigma(x_2)$ and $E(x_2)$ in the direct numerical simulation (dashed curves) and the theoretical approximation (solid curves). The time evolutions of $\sigma(x_2)$ and $E(x_2)$ in the direct numerical simulation are qualitatively similar to the theoretical approximation. Figure 4(c) shows a comparison of the time evolutions of $x_2(t)$ in the direct numerical simulation (dashed curve) and the theoretical approximation (solid curve).  
The period of the stepped motion in the direct numerical simulation is the same order but slightly larger than the theoretical model. We think that the main reason for the deviation is the rough approximation obtained using eqs.~(15) and (17) for the derivative of the electric potential at $x=x_2$.  

\section{Summary}
We have proposed a one-dimensional model for stepped leaders and a simplified dynamical model for the position $x_2$ of the stepped leader to understand the stepwise motion. It was shown that the two-step function of the conductance and the shift of the threshold by the form $E_{c2}=E_{c20}-\alpha\sigma$ is essentially important for the stepwise motion in our model. 
Our model might be very simple for the realistic discharge process, but we think that such a simplified model is useful for understanding the unique motion of the stepped leader. Stepwise growth was observed in other systems, such as bacteria colonies.~\cite{rf:6} Some similar mechanisms might work also in these systems.

\end{document}